\documentclass[11pt,preprint]{aastex}
\usepackage{natbib}
\usepackage{amsmath}
\begin{document}

\title{Atomic data for \ion{Zn}{2} - Improving Spectral Diagnostics of Chemical 
Evolution in High-redshift Galaxies}
\author{
Romas Kisielius\altaffilmark{1}, 
Varsha P. Kulkarni\altaffilmark{2}, 
Gary J. Ferland\altaffilmark{3,4},
Pavel Bogdanovich\altaffilmark{1}, 
Debopam Som\altaffilmark{2},
Matt L. Lykins\altaffilmark{3}
}

\altaffiltext{1}{Institute of Theoretical Physics and Astronomy, Vilnius 
University, A. Go{\v s}tauto 12, LT-01108, Lithuania}
\altaffiltext{2}{Department of Physics and Astronomy, University of South 
Carolina, Columbia, SC 29208, USA}
\altaffiltext{3}{Department of Physics and Astronomy, University of Kentucky, Lexington, KY 40506, USA}
\altaffiltext{4}{School of Mathematics and Physics, Queen's University Belfast, Belfast BT7 1NN, 
Northern Ireland, UK}

\begin{abstract}

Damped Lyman-alpha (DLA) and sub-DLA absorbers in quasar spectra provide the 
most sensitive tools for measuring element abundances of distant galaxies. 
Estimation of abundances from absorption lines depends sensitively on the 
accuracy of the atomic data used. We have started a project to produce new 
atomic spectroscopic parameters for optical/UV spectral lines using 
state-of-the-art computer codes employing very broad configuration interaction 
basis. Here we report our results for \ion{Zn}{2}, an ion used widely in studies 
of the interstellar medium (ISM) as well as DLA/sub-DLAs. We report new 
calculations of many energy levels of \ion{Zn}{2}, and the line 
strengths of the resulting radiative transitions. Our calculations use  
the configuration interaction approach within a numerical Hartree-Fock framework. 
We use both non-relativistic and quasi-relativistic one-electron radial 
orbitals. We have incorporated the results of these atomic calculations 
into the plasma simulation code Cloudy, and applied them to a lab plasma and examples 
of a DLA and a sub-DLA. Our values of the \ion{Zn}{2}
$\lambda \lambda$ 2026, 2062 oscillator strengths are higher than previous 
values by 0.10 dex. Cloudy calculations for representative absorbers with the 
revised Zn atomic data imply ionization corrections lower than calculated before 
by 0.05 dex. The new results imply Zn metallicities should be lower by 0.1 dex 
for DLAs and by 0.13-0.15 dex for sub-DLAs than in past studies. Our results can 
be applied to other studies of \ion{Zn}{2} in the Galactic and extragalactic ISM. 
\end{abstract}


\section{Introduction}

Most elements heavier than He are produced by stellar evolution and then 
distributed into interstellar space. Understanding the chemical composition of 
distant galaxies is therefore crucial to understanding the star formation and 
feedback processes central to galaxy evolution. Absorption lines of DLA and 
sub-DLA absorbers in the spectra of background quasars  provide the most 
sensitive tools to measure the heavy element content of distant galaxies. 
The DLAs have neutral hydrogen column densities 
$N_{\rm H I}  \ge 2 \times 10^{20}$ cm$^{-2}$, and the sub-DLAs have 
$10^{19} \le N_{\rm H I} < 2 \times 10^{20}$ cm$^{-2}$. The DLAs and sub-DLAs 
dominate the neutral gas content of galaxies, and constitute most of the H~I 
available for star formation at redshifts $0 <z < 5$ [e.g., \citet{Storrie00};
\citet{Per05}; \citet{Pro05}; \citet{Noter12}; \citet{Zafar13}].
DLAs observed toward GRB afterglows also offer an excellent 
probe of the physical and chemical conditions in distant galaxies [e.g., \citet{Savaglio03}; \citet{Chen05}; 
\citet{Pro07}; \citet{Fynbo09}].

Besides neutral hydrogen, DLAs and sub-DLAs also show a number of other elements 
ranging from C to Zn. The abundances of these elements provide very sensitive 
indicators of the chemical evolution of galaxies  [e.g., \citet{Pettini97}; 
\citet{KF02}; \citet{Pro03};  
\citet{Per08}; \citet{Mei09b}; \citet{Cooke11}; \citet{Rafelski12}; 
\citet{Som13, Som14}]. Since different elements are produced by stars of 
different masses, the measurements of  element abundances as a function of time 
give information about the history of formation of stars of different masses 
in galaxies. 

The quality of the atomic data directly affect the accuracy of the element abundances and 
physical properties of galaxies that are estimated from the measurements of absorption lines. 
The most commonly used atomic data reference for the analysis of absorption lines in DLAs and sub-DLAs is 
\cite{Morton03} [see, e.g., \citet{Battisti12}; \citet{Rafelski12}; 
\citet{Kulk12};  \citet{Guimaraes12}; \citet{Jorgenson13}; 
\citet{Som13, Som14}]. On the other hand, there is a need to improve beyond the 
oscillator strengths of \cite{Morton03}. This is because Morton (2003) lists large uncertainties for the oscillator strengths of some important transitions.  [These uncertainty values are listed on the NIST Atomic Spectra Database 
\citep{asd13}]. Furthermore, for some transitions, \cite{Morton03} lists no 
oscillator strengths at all. In some cases, the NIST database assigns low 
accuracy grades even for more recent values obtained since \cite{Morton03}. 
Such shortcomings in atomic data limit our ability to interpret the 
spectra of high-redshift galaxies, potentially leading to erroneous inferences 
about their chemical enrichment and star formation history. 

With the goal of producing new reliable atomic data for commonly used 
astrophysical ions, we have started a collaborative study combining  
atomic physics, plasma simulations, and observational spectroscopy. The goals 
of this study are to examine the quality of available atomic data, to improve 
the accuracy of the atomic data with low reported accuracies, to incorporate these new atomic data 
into Cloudy (our widely utilized plasma simulation code), and to study the effect of the revised Cloudy code on the analysis of absorption lines in DLAs and sub-DLAs. In a recent study (Kisielius et al. 2014), we examined the atomic data for the 
key ion \ion{S}{2}. Here we focus on the ion \ion{Zn}{2}, which also plays 
a very important role in studies of DLAs and sub-DLAs.

\subsection{Why \ion{Zn}{2}?}

Refractory elements such as Fe, Si, or Mg condense in the form of solid dust 
grains in the interstellar medium. On the other hand, volatile elements such as 
N, O, P, S, Ar, Zn do not condense appreciably on interstellar dust grains. 
The gas-phase abundances of such weakly depleted elements can therefore give 
their total (gas + solid phase) abundances. For observations of DLAs and 
sub-DLAs, weak, unsaturated lines of the elements N, O, P, or Ar are often not 
accessible in ground-based spectroscopy, which makes it difficult to measure 
their column densities reliably. Moreover, the abundance of N appears to be 
complicated by nucleosynthetic differences between primary and secondary N 
production [e.g., \citet{Pet95}; \citet{Som14}]. The two elements that 
have emerged as the most useful for DLA/sub-DLA studies are S and Zn. Having 
discussed the atomic data for S in \cite{rk2014}, we now turn to Zn.

Zn is especially interesting because it tracks Fe closely in Galactic stars 
(for [Fe/H] $\gtrsim -2$). Especially important among the Zn ions is \ion{Zn}{2}, 
which is the dominant ion in DLAs. A key advantage of  \ion{Zn}{2} is that it 
has two weak absorption lines in a relatively narrow wavelength region at 
$\lambda \, \lambda 2026.14, 2062.66$, that are often unsaturated and hence allow 
accurate column density determinations. Moreover, these lines often lie outside 
the Lyman-$\alpha$ forest, and therefore allow unambiguous identifications and 
measurements free of blends. This has made Zn the most commonly used metallicity 
indicator for DLAs and sub-DLAs. Starting from the early work of 
\cite{Mey89} and \cite{Pet94}, Zn has been observed in $> 150$ 
DLAs and $\sim 50$ sub-DLAs at redshifts ranging from $z < 0.1$ to $z > 3.3$. 
 Zn has also been used as a metallicity indicator in studies of the Milky Way interstellar gas. 
For all of these reasons, accurate atomic data for \ion{Zn}{2} transitions are very important. 

\cite{Morton03} lists the oscillator strengths of 0.501 and 0.246 for the  
\ion{Zn}{2} $\lambda \, \lambda \, 2026.14, 2062.66$ lines, respectively, but there are no 
estimates of the  uncertainties in these values in the NIST database. 
Thus, the uncertainties in the metallicity introduced by the uncertainty in 
the oscillator strengths could be far larger than those often 
quoted from the measurement uncertainties in high-resolution data (typically 
$\lesssim 0.05$ dex). \ion{Zn}{2} has additional absorption lines at 923.98, 938.71, 
949.46, 984.14, and 986.52 {\AA}, but they are listed in Morton (2003) without 
any oscillator strength estimates.

\subsection{Previous Zn II Calculations}
\label{previous_z2}

There is quite a generous amount of either experimental or theoretical studies
considering the transition wavelengths, radiative transitions or scattering 
processes in the ion \ion{Zn}{2}, see, e.g., experimental works of 
\citet{berlaw93, mayo2006, gull00}.
Multiconfiguration Hartree-Fock calculations were reported by \citet{cff77}.
The Hartree-Fock approximation adopting transformed radial orbitals was used
by \citet{rkpb01_zn} to calculate oscillator strengths of astrophysically 
important lines in \ion{Zn}{1} and \ion{Zn}{2} ions. Their calculations which
included core-polarization effects produced oscillator strengths which agreed 
quite well with previously published semiempirical values.
Recently \citet{hh2003} have presented a pivotal study of oscillator strengths 
for the \ion{Zn}{2}. They have investigated the 4s--4p resonance line oscillator 
strengths in the \ion{Zn}{2} ion using extensive configuration interaction (CI) 
calculations. They studied the influence of core-polarization, 
electron-correlation in the core, core-core correlation effects in resulting
oscillator strengths using CIV3 computer code which deals with relativistic 
effects in Breit-Pauli approximation. Their determined oscillator strength 
values, $f=0.268$ for the $4s\, ^2S_{1/2} - 4p\, ^2P^{\mathrm{o}}_{1/2}$ line 
and $f=0.547$ for the $4s\, ^2S_{1/2} - 4p\, ^2P^{\mathrm{o}}_{3/2}$ line lie
about $5-10 \%$ higher than the recent experimental values obtained by
\citet{berlaw93}, $f=0.255 \pm 0.024$ and $f=0.492 \pm .039$, respectively. 
Nevertheless, the theoretical $f$ values of \citet{hh2003} are in good agreement 
with relativistic many-body perturbation theory calculation results of 
\citet{chou97}, which are $f=0.264$ and $f=0.538$ for these lines. Unfortunately,
\citet{hh2003} have considered only 4s -- 4p lines, therefore their data 
are not enough for a comprehensive modeling calculations and can not be employed
in our investigation.

Very recently \citet{celik2013} published new calculations of atomic data for 
the \ion{Zn}{2} ion. Their calculations were performed using two different
semi-empirical methods, the weakest bound electron potential model theory 
(WBEPMT)and the quantum defect orbital theory (QDOT). They employed numerical 
Coulomb approximation wave functions and numerical non-relativistic Hartree-Fock 
wave functions to determine the necessary parameters. As a result, the real 
multi-particle system was transformed to a simple one-particle system. In the
WBEPMT case, the effective parameters (nuclear charge $Z^*$, $n^*$, $l^*$) were 
derived from the experimental energy data or from other existing calculations 
for the mean radius $<r>$. One can have a reasonable doubt if these $<r>$
values, derived from the different sources, are really coherent and consistent.
Unfortunately, their suitability for the 4s electrons is very questionable as 
their wavefunctions significantly overlap with those of other electrons. 
As a result, the results from these two different approximations differ quite 
significantly one from another even if they both are semi-empirical and based 
on the same experimental data.

Several works have considered electron-impact excitation parameters for the
\ion{Zn}{2} ion. Scattering parameters for some transitions in this ion were 
determined by \citet{pind91, oleg99} by employing R-matrix methods in 
the $LS$-coupling approximation. \citet{sharma11} have applied a fully 
relativistic distorted-wave theory to study the electron-impact excitation of 
the $ns - np$ resonance transitions in singly-charged metal ions with one 
valence electron, including Zn$^+$ ions. Unfortunately, the above-mentioned studies
did not produce atomic data sets suitable for spectral modeling where the
complete and consistent data are required.

In order to assess the accuracy of the Zn II atomic data, we performed new 
calculations of the oscillator strengths for all \ion{Zn}{2} electric dipole, 
magnetic dipole, and electric quadrupole transitions. Section 2 describes our  
new calculations and how they compare with previous estimates. Section 3 outlines  
the inclusion of these calculations into Cloudy and gives a few examples of applications. Finally, section 4 summarizes our results and their implications for abundance studies of DLAs and sub-DLAs.

\section{Calculations of new atomic data}
\label{method}

Our calculations are performed by employing Hartree-Fock radial orbitals (HFRO).
The relativistic corrections are included in the Breit-Pauli approximation. 
We determine spectral parameters for four even configurations $3d^{10}4s$, 
$3d^94s^2$, $3d^{10}4d$ ,and $3d^{10}5s$ and for three odd configurations 
$3d^{10}4p$, $3d^{10}5p$, and $3d^94s4p$. The configuration $3d^94s4p$ levels 
lie in an energy range which is significantly wider compared to the purpose 
of this work. For that reason we determine only the lowest levels of this 
configuration arising from the $4s$ and $4p$ electrons bound into $^1P$ term. 
The electron-correlation effects are included in the configuration 
interaction (CI) approximation by adopting the basis of transformed radial 
orbitals (TRO) described by \citet{pb99a}. 

At the first step, we solve the HF equations for the ground configuration using 
the code of \citet{cff87}. Next, we determine solutions of HF equations 
$P(nl|r)$ for all $4l$, $5s$ and $5p$ orbitals in a frozen-core potential.  
This basis of HFRO is complemented with transformed radial orbitals  
$P_{\mathrm{TRO}}(nl|r)$, which are introduced to describe virtual electron 
excitations from adjusted (investigated) configurations. The TRO are obtained 
by a way of transformation:

\begin{align}
\label{tro}
P_{\mathrm{TRO}}(nl|r) = &
N ( r^{l-l_0+k} \exp(-Br)P(n_0l_0|r) \nonumber \\ 
- & \sum_{n^{\prime} < n} P(n^{\prime}l|r) \int_{0}^{\infty} 
P(n^{\prime}l|r^{\prime})r^{\prime\;(l-l_0+k)} \exp(-Br^{\prime})
P(n_0l_0|r^{\prime})dr^{\prime} ) .
\end{align}

The parameters $k$ and $B$ are introduced to ensure the maximum of the  mean 
energy corrections determined in the second-order perturbation theory (PT)
(see \cite{pb99a}). Here the factor $N$ ensures the normalization of the 
TROs, which are determined for the electrons with the principal quantum number 
values $6 \leq n \leq 11$ and for all allowed values of the orbital quantum 
number $l$.

The set of admixed configurations is generated from the adjusted configurations 
by introducing one-electron and two-electron excitations from the 3p, 3d, 4s, 
4p, 4d, and 5s shells to all available states in the basis of determined radial 
orbitals. This leads to a huge set of admixed configurations and, consequently, 
to large Hamiltonian matrices. In order to reduce the size of the Hamiltonian 
matrices to be diagonalized, we need to determine the  most significant 
configurations. As a selection criterion, we use the averaged weights ${\bar W}$
of the admixed configurations $K^{\prime}T^{\prime}LS$ in the CI wavefunction 
expansion of the adjusted configuration $K_0TLS$: 
\begin{equation}
\label{eq-w}
{\bar W}_{\mathrm{PT}}(K_0,K^{\prime}) = 
\frac{\sum_{TLST^{\prime}}(2L+1)(2S+1) 
\langle K_0TLS \Vert H \Vert K^{\prime}T^{\prime}LS \rangle ^2}
{g(K_0) \left( {\bar E}(K^{\prime}) - {\bar E}(K_0) \right)^2},
\end{equation}
where $T$ describes all possible intermediate momenta, which bound the 
non-relativistic configurations $K$ and $K_0$ into $LS$ multiplet.
These averaged weights are determined in the second-order PT. We select only 
those configurations which have their weights larger than some selection
criterion $W_0$. More details on configuration selection procedure are given by
\citet{pbrk01}.

Bearing in mind restrictions of our computer system, we can perform our 
spectroscopic data generation for $W_0 \approx 5 \times 10^{-6}$. The main 
limiting factor is the size $M$ of the electrostatic interaction operator $H_0$ 
which depends on the number of configuration state functions (CSF) having the 
same total momenta $LS$. We have performed two other similar test calculations 
with the value of $W_0$ increased by a factor of two 
(CI$_{\mathrm{TRO}}^{\mathrm{red1}}$) and by a factor of ten 
(CI$_{\mathrm{TRO}}^{\mathrm{red2}}$). In the further description we present 
the parameters of these three calculations separated by a slash. 
In that way we can demonstrate the convergence of our  calculated results.

For the four even configurations investigated, we have selected the most 
important $S_\mathrm{e} = 1012/745/338$ non-relativistic configurations, 
including adjusted ones, which give rise to $C_\mathrm{e} = 749470/517350/209921$ 
configuration state functions (CSF). Here we apply CSF
reduction by moving the shells of virtually excited electrons to the beginning 
of active shells, as it was described in detail by \citet{pbrkam02, pbam04}. 
This procedure reduces the number of CSFs to  $R_\mathrm{e} = 20991/16021/7522$.
Using this basis, the size of the largest $H_0$ matrix is  
$M_\mathrm{e} = 17317/13195/6131$. For the three odd configurations investigated,
we have selected $S_\mathrm{o} = 784/577/252$ most important configurations, 
including the adjusted ones. These configurations make 
$C_\mathrm{o} = 2513717/1727977/721017$ CSFs, which in turn are further reduced by 
the CSF-reduction procedure to $R_\mathrm{o} = 531815/386459/181535$. Adopting this
base, the size of largest $H_0$ matrix reaches $M_\mathrm{o} = 108720/78903/36696$.

The Hamiltonian eigenvalues and eigenfunctions are determined adopting this 
reduced configuration basis. Further, the determined CI wavefunctions are 
employed to calculate electron transition parameters. The M1 and E2 radiative 
transition data are produced for transitions among the levels of the same-parity 
configurations, and the E1, M2, and E3 transition parameters are produced for 
the radiative transitions among the levels of the different-parity 
configurations. The significance of the radiative transitions of higher 
multipole order, such as M2 or E3, for the radiative lifetimes of some levels 
was demonstrated by \citet{rasa13}.

Following the methods of \citet{pbrkds14}, we also produce electron-impact 
excitation collision strengths in the plane-wave Born approximation. The most 
inclusive description of the adopted approach is given by \citet{tro04, tro05}, 
whereas its application for data production and their accuracy analysis is 
given by \citet{pbim99, pbrkim03, pbrkau03, rkpb03, ak06}.

To perform our calculations, we have employed our own computer codes together 
with the codes of \citet{ahrgcff91, cffmgah91, cffmg91} which have been adopted 
for our computing needs. The code of \citet{ahrgcff91} has been updated 
according to the methods presented by \citet{gg97, gg98}.

\subsection{Energy levels and wavelengths}

\renewcommand{\baselinestretch}{0.85}
\begin{deluxetable}{rlrrrrrrrr}
\tabletypesize{\footnotesize}
\tablecolumns{8} 
\tablewidth{0pt}
\tablecaption{
\label{tab_lev}
Comparison of calculated \ion{Zn}{2} level energies $E$ (in cm$^{-1}$) 
and their percentage deviations $\delta E$ with experimental data 
from the NIST database. 
} 
\tablehead{ 
\colhead{$N$} &
\colhead{State} & 
\colhead{$2J+1$} &
\colhead{NIST} & 
\colhead{CI$_{\mathrm{TRO}}$} & 
\colhead{$\delta E_{\mathrm{TRO}}$} & 
\colhead{CI$_{\mathrm{TRO}}^{\mathrm{red1}}$} & 
\colhead{$\delta E_{\mathrm{TRO}}^{\mathrm{red1}}$} &
\colhead{CI$_{\mathrm{TRO}}^{\mathrm{red2}}$} & 
\colhead{$\delta E_{\mathrm{TRO}}^{\mathrm{red2}}$} \\
}
\startdata 
  1&  $3d^{10}4s       $\;$^2S$&  2&      0.00&      0&    - &         0&       &     0&   -  \\
  2&  $3d^{10}4p       $\;$^2P$&  2&  48481.00&  48670&  0.39&     47530& -1.96 & 48546& 0.13 \\
  3&  $3d^{10}4p       $\;$^2P$&  4&  49355.04&  49144& -0.43&     48004& -2.74 & 49018&-0.68 \\
  4&  $3d^{9}4s^2      $\;$^2D$&  6&  62722.45&  62815&  0.15&     63515&  1.26 & 66605& 6.19 \\
  5&  $3d^{9}4s^2      $\;$^2D$&  4&  65441.64&  65404& -0.16&     66110&  1.02 & 69292& 5.88 \\
  6&  $3d^{10}5s       $\;$^2S$&  2&  88437.15&  88394& -0.05&     88059& -0.43 & 92729& 4.85 \\
  7&  $3d^{10}4d       $\;$^2D$&  4&  96909.74&  96972& -0.06&     96864& -0.05 &102361& 5.62 \\
  8&  $3d^{10}4d       $\;$^2D$&  6&  96960.40&  96988&  0.03&     96880& -0.08 &102377& 5.59 \\
  9&  $3d^{10}5p       $\;$^2P$&  2& 101365.9 & 101571&  0.20&    100818& -0.54 &106727& 5.29 \\
 10&  $3d^{10}5p       $\;$^2P$&  4& 101611.4 & 101705&  0.09&    100953& -0.65 &106865& 5.17 \\
 11&  $3d^{9}(^2D)4s4p $\;$^4P$&  6& 103701.6 & 105603&  1.83&    106286&  2.49 &108677& 4.80 \\	
 12&  $3d^{9}(^2D)4s4p $\;$^4P$&  4& 105322.7 & 106855&  1.45&    107526&  2.09 &109935& 4.38 \\	
 13&  $3d^{9}(^2D)4s4p $\;$^4P$&  2& 106528.8 & 107849&  1.24&    108530&  1.88 &110936& 4.14 \\	
 14&  $3d^{9}(^2D)4s4p $\;$^4F$& 10& 106779.9 & 107465&  0.64&    108014&  1.16 &110474& 3.46 \\  
 15&  $3d^{9}(^2D)4s4p $\;$^4F$&  8& 106852.4 & 107938&  1.03&    108470&  1.51 &110940& 3.82 \\	
 16&  $3d^{9}(^2D)4s4p $\;$^4F$&  6& 107268.6 & 108515&  1.16&    109039&  1.65 &111522& 3.96 \\
 17&  $3d^{9}(^2D)4s4p $\;$^4F$&  4& 108227.9 & 109435&  1.11&    109973&  1.61 &112437& 3.89 \\	
 18&  $3d^{9}(^2D)4s4p $\;$^2F$&  6& 110672.3 & 111988&  1.19&    112402&  1.56 &114951& 3.87 \\	
 19&  $3d^{9}(^2D)4s4p $\;$^4D$&  8& 110867.2 & 110683& -0.17&    111004&  0.12 &113435& 2.32 \\	
 20&  $3d^{9}(^2D)4s4p $\;$^4D$&  6& 111743.0 & 111093& -0.58&    111453& -0.26 &113950& 1.97 \\	
 21&  $3d^{9}(^2D)4s4p $\;$^4D$&  4& 111994.3 & 112044&  0.04&    112375&  0.34 &114833& 2.53 \\	
 22&  $3d^{9}(^2D)4s4p $\;$^2F$&  8& 112409.7 & 112763&  0.31&    113204&  0.71 &115809& 2.91 \\	
 23&  $3d^{9}(^2D)4s4p $\;$^4D$&  2& 112534.9 & 113409&  0.78&    113702&  1.04 &115135& 2.31 \\	
 24&  $3d^{9}(^2D)4s4p $\;$^2P$&  2& 113492.9 & 112272& -1.08&    112571& -0.81 &116293& 2.46 \\	
 25&  $3d^{9}(^2D)4s4p $\;$^2P$&  4& 113499.2 & 113479& -0.02&    113768&  0.24 &116476& 2.62 \\	
 26&  $3d^{9}(^2D)4s4p $\;$^2D$&  4& 114045.03& 114126&  0.07&    114440&  0.35 &117017& 2.61 \\	
 27&  $3d^{9}(^2D)4s4p $\;$^2D$&  6& 114833.95& 114759& -0.07&    115085&  0.22 &117659& 2.46 \\	
 \tableline
$MSD$&                         &   &          &    730 &      &      1146 &       &  3809 &      \\  
\enddata                
                        
\tablecomments{        
NIST - experimental energies from NIST database;
CI$_{\mathrm{TRO}}$ - our HF data with a complete CI expansion using TRO;
CI$_{\mathrm{TRO}}^{\mathrm{red1}}$ - our HF data from the reduced CI expansion 
calculation; 
CI$_{\mathrm{TRO}}^{\mathrm{red2}}$ - our HF data from the reduced CI expansion 
calculation; 
$\delta E_{\mathrm{TRO}}$ - percentage deviations of HF data from the 
experimental energies;
$\delta E_{\mathrm{TRO}}^{\mathrm{red1}}$ - percentage deviations of the reduced 
CI data from the observed energies;
$\delta E_{\mathrm{TRO}}^{\mathrm{red2}}$ - percentage deviations of the reduced 
CI data from the observed energies.}
\end{deluxetable}

We present a comparison of our calculated \ion{Zn}{2} level energies with the 
data extracted from the NIST database \citep{asd13} in Table~\ref{tab_lev}.
As we have explained in Sect.~\ref{method}, three different CI expansions are
adopted in the \ion{Zn}{2} level energies calculation using the same 
multiconfiguration Hartree-Fock method based on the TRO. As one can clearly see,
the agreement of the data becomes closer when the selection criterion 
${\bar W}_{\mathrm{PT}}$ goes down. The mean-square deviations $MSD$ are 
given in the last row of Table~\ref{tab_lev} for the indication of the 
convergence of our calculation. One can notice a considerable improvement of 
accuracies of the calculated energies when the selection criterion 
${\bar W}_{\mathrm{PT}}$ is decreased two times (from $MSD=3809$ to $MSD=1146$\;
cm$^{-1}$), and a smaller improvement when this criterion is further decreased 
five times (down to $MSD=730$\;cm$^{-1}$). 

Alongside the level energies determined using the different CI expansions, we 
present the corresponding percentage deviations $\delta E$:
\begin{equation}
\delta E= \frac{E_{\mathrm{TRO}}-E_{\mathrm{NIST}}}{E_{\mathrm{NIST}}} \cdot 100 \%.
\end{equation}
As one can notice, the $\delta E$ decrease consistently for most levels, as 
the CI expansion increases. This again is underlining the fact that (i) the
accuracy of our energy levels increases when the extended CI basis is employed; 
and (ii) the convergence for the calculated level energies is achieved as the 
final changes in the $\delta E$ values are much smaller compared to the initial 
ones. Based on our past experience with such calculations, we do not expect 
further extensions of the CI basis to yield any substantial changes of the 
level energies or the radiative transition parameters.

Here we should explain that in further consideration or in modeling performed by 
Cloudy we do not use the calculated energies. They are substituted by 
the experimental values in order to enable us to have the correct transition 
wavelengths. The main  reason to present Table~\ref{tab_lev} here is to 
demonstrate the convergence of our calculations for the energy levels, which 
reflects on the production of transition parameters such as oscillator 
strengths $gf$ or transition probabilities $A$.

\begin{figure}
\includegraphics[scale=0.8]{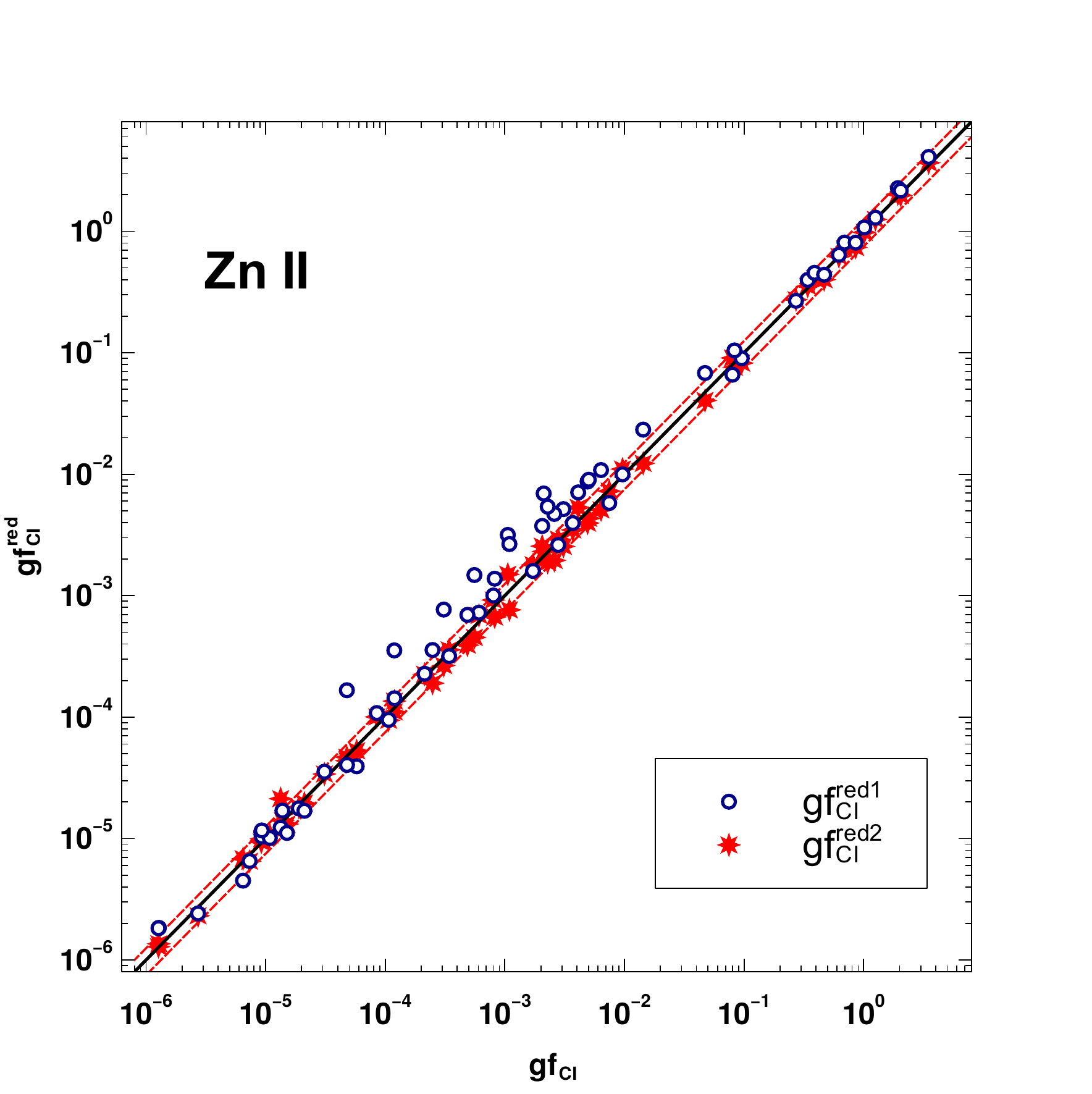}
\caption{ 
Comparison of the weighted oscillator strengths $gf$ determined in the 
CI$_{\mathrm{TRO}}$ to those determined in CI$_{\mathrm{TRO}}^{\mathrm{red1}}$ 
and CI$_{\mathrm{TRO}}^{\mathrm{red2}}$ approaches for \ion{Zn}{2}.
Dashed lines show the 25\% deviation limits.
\label{oschf1}
}
\end{figure}

Using the determined CI wavefunction expansions, we are able to generate data 
sets for the radiative transitions involving twenty seven energy levels. 
In Fig.~\ref{oschf1} we compare the weighted oscillator strengths determined
in approximation CI$_{\mathrm{TRO}}$ with $gf$ values determined in
CI$_{\mathrm{TRO}}^{\mathrm{red1}}$ and  CI$_{\mathrm{TRO}}^{\mathrm{red2}}$
approximations. Although there are some weaker lines where deviations can exceed 
$25 \%$, especially in CI$_{\mathrm{TRO}}^{\mathrm{red2}}$ approximation, 
for most lines the agreement is within the range of $25 \%$.
One can notice that the deviations have become significantly smaller in
CI$_{\mathrm{TRO}}^{\mathrm{red1}}$ approximation, indicating the convergence of
the $gf$ values, as it has been a case with level energy values (see 
Table~\ref{tab_lev}).

\renewcommand{\baselinestretch}{0.9}
\begin{deluxetable}{llrrl}
\tabletypesize{\footnotesize}
\tablecolumns{5} 
\tablewidth{0pt}
\tablecaption{
\label{tab_tran}
Transition line strengths $S$ (in a.u.) for \ion{Zn}{2} determined in the 
CI$_{\mathrm{TRO}}$ approximation. 
} 
\tablehead{ 
\colhead{Data} &
\colhead{Type} & 
\colhead{$N_l$} &
\colhead{$N_u$} &
\colhead{$S$}
}
\startdata 
S& E1&  1&   2& 4.199E$+$00\\
S& E1&  1&   3& 8.401E$+$00\\
S& M2&  1&   3& 1.405E$+$02\\
S& E2&  1&   4& 1.525E$+$00\\
S& E2&  1&   5& 9.898E$-$01\\
S& M1&  1&   6& 1.115E$-$06\\
S& E2&  1&   7& 6.011E$+$01\\
S& E2&  1&   8& 9.013E$+$01\\
S& E1&  1&   9& 1.001E$-$02\\
S& E1&  1&  10& 1.593E$-$02\\
S& M2&  1&  10& 2.201E$-$02\\
S& E3&  1&  11& 9.030E$-$06\\
S& M2&  1&  11& 1.250E$+$01\\
S& E1&  1&  12& 2.970E$-$02\\
S& M2&  1&  12& 2.061E$+$00\\
S& E2&  2&   3& 1.278E$+$02\\
S& M1&  2&   3& 1.300E$+$00\\
S& E3&  2&   4& 3.139E$+$00\\
S& M2&  2&   4& 8.302E$-$03\\
S& E1&  2&   5& 1.196E$-$02\\
S& M2&  2&   5& 7.790E$-$03\\
S& E1&  2&   6& 2.822E$+$00\\
S& E1&  2&   7& 1.317E$+$01\\
S& M2&  2&   7& 8.199E$+$00\\
\enddata      
              
\tablecomments {
The first column describes the transition data type (S: line strength $S$, 
A: transition probability $A$). 
The second column describes transition line type, 
$N_l$ is for the lower level index, 
$N_u$ denotes the upper level index.}
\tablecomments {
(This table is available in its entirety in a machine-readable form in the online
journal. A portion is shown here for guidance regarding its form and content.)
}
\end{deluxetable}

Table~\ref{tab_tran} gives a sample of the transition parameters used in the 
plasma simulation package Cloudy. Here transition line strengths $S$ are
tabulated together with the type of transition ($E1$, $E2$, $E3$, $M1$, $M2$). 
The line strengths $S$ are preferred as they do not depend explicitly on the
transition energy $\Delta E$. For the further use, oscillator strengths $f$
or transition probabilities $A$ can be derived using well-known relations.
A complete table of transition parameters is available on-line.

\renewcommand{\baselinestretch}{0.9}
\begin{deluxetable}{rlrrrrrr}
\tabletypesize{\footnotesize}
\tablecolumns{8} 
\tablewidth{0pt}
\tablecaption{
\label{tab_gf}
Transition vacuum wavelengths (in \AA), the upper  levels, oscillator strengths $gf$ 
and transition probabilities $A$ (in s$^{-1}$) for \ion{Zn}{2}, including some 
commonly observed lines. 
} 
\tablehead{ 
\colhead{$\lambda$ (\AA)} &
\colhead{Upper level} & 
\colhead{$gf$} &
\colhead{$A$} \\
}
\startdata 
 \tableline
2062.6604 &  $3d^{10}4p$\;$^2P_{1/2} $& 6.18E$-$1 & 4.85E$+$8 \\
2026.1370 &  $3d^{10}4p$\;$^2P_{3/2} $& 1.26E$+$0 & 5.12E$+$8 \\
 986.5237 &  $3d^{10}5p$\;$^2P_{1/2} $& 3.08E$-$3 & 1.06E$+$7 \\
 984.1414 &  $3d^{10}5p$\;$^2P_{3/2} $& 4.92E$-$3 & 8.46E$+$6 \\
 949.4630 &  $3d^{9}4s4p$\;$^4P_{3/2}$& 9.50E$-$3 & 1.76E$+$7 \\
 938.7130 &  $3d^{9}4s4p$\;$^4P_{1/2}$& 3.65E$-$3 & 1.38E$+$7 \\
 923.9760 &  $3d^{9}4s4p$\;$^4F_{3/2}$& 9.17E$-$6 & 1.79E$+$4 \\
\enddata                
\tablecomments {
All these lines originate from the ground level $3d^{10}4s$\;$^2S_{1/2}$ ($g = 2$).
The statistical weights $g = 2J+1$ of the upper levels are given 
in Table~\ref{tab_lev}.
}
\end{deluxetable}

In Table~\ref{tab_gf} we provide the transition wavelengths $\lambda$, 
oscillator strengths $gf$, and $A$ values for several observed lines. We note 
that for 5 of these lines, no $gf$ or $A$ values are listed in \citet{Morton03}.

Along with the radiative transition data, we have determined  electron-impact
excitation parameters for these lines in the plane-wave Born approximation.
The methods and codes for such calculation are described by
\citet{pbrkds14}. We present just a sample of the collisional parameters
in Table~\ref{tab_cs}, whereas the complete version of the table is available
on-line. We note that the approximation adopted in our calculations can not 
produce highly accurate data for the electron-impact excitation process.
Nevertheless, as the more accurate calculations for this level set of 
\ion{Zn}{2} are absent, our data are an improvement over other more rough 
approximations, such as g-bar approximation.

As already mentioned in Sect.~\ref{previous_z2}, several elaborate studies on 
the \ion{Zn}{2} ion electron-impact excitation were published, see 
\citet{pind91, oleg99, sharma11}. Since their data are incompatible with our 
model due to a different atomic structure, a narrow energy range (as in 
\citet{pind91, oleg99}), and a different data type, when only excitation cross 
sections are plotted instead of collision rates, we cannot include those 
data into our data set or perform a comprehensive comparison with our data. 
Some qualitative assesment suggests that our data deviate no more than $30\%$ 
from the R-matrix or relativistic distorted-wave results at low electron 
energies for the optically allowed transitions and no more than $45\%$ for the 
forbidden transitions. At high electron energies, these deviations decrease at 
least twofold.

The {\sl ab initio} calculation results for the level energies $E$, the 
radiative transition parameters - oscillator strengths $gf$, transition line 
strengths $S$, transition probabilities $S$, and the radiative lifetimes 
$\tau$ for the \ion{Zn}{2} are available from the database ADAMANT 
(\url{http://www.adamant.tfai.vu.lt/database}) 
being developed at Vilnius University.

 We note in passing that the theoretical calculations of the transition wavelengths  can never 
be as accurate as the experimental data. We have therefore used the experimental $\lambda$ values in our Cloudy simulation runs, as has been done in previous studies. The oscillator strengths or 
transition probabilities need to be corrected for the 
difference between the experimental and theoretical energies. 
We have indeed corrected the transition rates by using the experimental level 
energies to determine the radiative transition parameters (e.g., the transition probabilities $A$ or the oscillator strengths $gf$).

\begin{deluxetable}{llrrrrrrrrrrrrrr}
\tabletypesize{\scriptsize}
\tablecolumns{16} 
\tablewidth{0pt}
\tabcolsep 3pt
\rotate
\tablecaption{
Table 4: 
Effective collision strengths $\Upsilon$ for the electron-impact excitation 
of \ion{Zn}{2} at 14 selected temperatures determined in the plane-wave Born 
approximation.
\label{tab_cs}
} 
\tablehead{ 
&&
\multicolumn{14}{c}{Electron Temperatures (K)}\\
\cline{3-16}\\
\colhead{$N_l$}           &
\colhead{$N_{up}$}        & 
\colhead{$4 \times 10^2$} & 
\colhead{$8 \times 10^2$} & 
\colhead{$2 \times 10^3$} & 
\colhead{$4 \times 10^3$} & 
\colhead{$8 \times 10^3$} & 
\colhead{$2 \times 10^4$} & 
\colhead{$4 \times 10^4$} & 
\colhead{$8 \times 10^4$} & 
\colhead{$2 \times 10^5$} & 
\colhead{$4 \times 10^5$} & 
\colhead{$8 \times 10^5$} & 
\colhead{$2 \times 10^6$} & 
\colhead{$4 \times 10^6$} & 
\colhead{$8 \times 10^6$} \\ 
}
\startdata 
 1&  2& 
2.29E$+$0& 2.29E$+$0& 2.30E$+$0& 2.38E$+$0& 2.67E$+$0& 3.58E$+$0& 4.78E$+$0& 6.46E$+$0& 9.42E$+$0& 1.22E$+$1& 1.52E$+$1& 1.97E$+$1& 2.34E$+$1& 2.72E$+$1\\
 1&  3& 
4.53E$+$0& 4.53E$+$0& 4.55E$+$0& 4.70E$+$0& 5.26E$+$0& 7.07E$+$0& 9.45E$+$0& 1.28E$+$1& 1.87E$+$1& 2.41E$+$1& 3.03E$+$1& 3.92E$+$1& 4.65E$+$1& 5.41E$+$1\\
 1&  4& 
6.93E$-$2& 6.93E$-$2& 6.94E$-$2& 7.05E$-$2& 7.57E$-$2& 9.37E$-$2& 1.15E$-$1& 1.40E$-$1& 1.67E$-$1& 1.82E$-$1& 1.92E$-$1& 1.99E$-$1& 2.02E$-$1& 2.03E$-$1\\
 1&  5& 
4.67E$-$2& 4.67E$-$2& 4.67E$-$2& 4.74E$-$2& 5.06E$-$2& 6.22E$-$2& 7.64E$-$2& 9.23E$-$2& 1.10E$-$1& 1.20E$-$1& 1.26E$-$1& 1.31E$-$1& 1.33E$-$1& 1.35E$-$1\\
 1&  6& 
3.17E$-$1& 3.17E$-$1& 3.17E$-$1& 3.19E$-$1& 3.33E$-$1& 4.07E$-$1& 5.22E$-$1& 6.84E$-$1& 9.39E$-$1& 1.13E$+$0& 1.29E$+$0& 1.45E$+$0& 1.54E$+$0& 1.59E$+$0\\
 1&  7& 
3.13E$-$1& 3.13E$-$1& 3.13E$-$1& 3.14E$-$1& 3.25E$-$1& 3.85E$-$1& 4.81E$-$1& 6.20E$-$1& 8.41E$-$1& 1.01E$+$0& 1.16E$+$0& 1.30E$+$0& 1.38E$+$0& 1.43E$+$0\\
 1&  8& 
4.68E$-$1& 4.68E$-$1& 4.68E$-$1& 4.70E$-$1& 4.86E$-$1& 5.76E$-$1& 7.21E$-$1& 9.29E$-$1& 1.26E$+$0& 1.51E$+$0& 1.73E$+$0& 1.95E$+$0& 2.06E$+$0& 2.14E$+$0\\
 1&  9& 
1.02E$-$1& 1.02E$-$1& 1.02E$-$1& 1.02E$-$1& 1.04E$-$1& 1.16E$-$1& 1.32E$-$1& 1.51E$-$1& 1.74E$-$1& 1.86E$-$1& 1.93E$-$1& 2.00E$-$1& 2.05E$-$1& 2.11E$-$1\\
 2&  3& 
1.76E$-$1& 2.85E$-$1& 5.68E$-$1& 9.45E$-$1& 1.48E$+$0& 2.35E$+$0& 2.99E$+$0& 3.48E$+$0& 3.87E$+$0& 4.02E$+$0& 4.11E$+$0& 4.16E$+$0& 4.18E$+$0& 4.19E$+$0\\
 2&  4& 
8.77E$-$7& 8.87E$-$7& 9.69E$-$7& 1.13E$-$6& 1.36E$-$6& 1.72E$-$6& 2.03E$-$6& 2.37E$-$6& 2.82E$-$6& 3.09E$-$6& 3.28E$-$6& 3.41E$-$6& 3.46E$-$6& 3.49E$-$6\\
 2&  5& 
1.15E$-$5& 1.16E$-$5& 1.27E$-$5& 1.54E$-$5& 2.07E$-$5& 3.21E$-$5& 4.27E$-$5& 5.27E$-$5& 6.27E$-$5& 7.41E$-$5& 2.86E$-$4& 2.61E$-$3& 7.28E$-$3& 1.44E$-$2\\
 2&  6& 
2.67E$-$1& 2.67E$-$1& 2.72E$-$1& 3.06E$-$1& 4.25E$-$1& 8.37E$-$1& 1.45E$+$0& 2.40E$+$0& 4.26E$+$0& 6.10E$+$0& 8.23E$+$0& 1.14E$+$1& 1.41E$+$1& 1.67E$+$1\\
 2&  7& 
2.21E$-$5& 2.21E$-$5& 2.25E$-$5& 2.69E$-$5& 4.83E$-$5& 1.55E$-$4& 3.74E$-$4& 8.11E$-$4& 1.88E$-$3& 3.12E$-$3& 5.07E$-$3& 1.41E$-$1& 1.28E$+$0& 4.88E$+$0\\
\enddata 
\tablecomments {
(This table is available in its entirety in a machine-readable form in the online
journal. A portion is shown here for guidance regarding its form and content.)
}
\end{deluxetable} 

\clearpage

\section{Cloudy calculations}
\label{cloudy}


\subsection{Application to DLAs and Sub-DLAs}
\label{dla}

One of the motivations behind our interest in the \ion{Zn}{2} atomic data stems 
from the somewhat surprising results obtained from observations of \ion{Zn}{2} 
lines in DLAs and sub-DLAs. We now describe these results and discuss the 
implications of our revised atomic data for the evolution of DLAs/ sub-DLAs.

\begin{figure}
\includegraphics[scale=0.67]{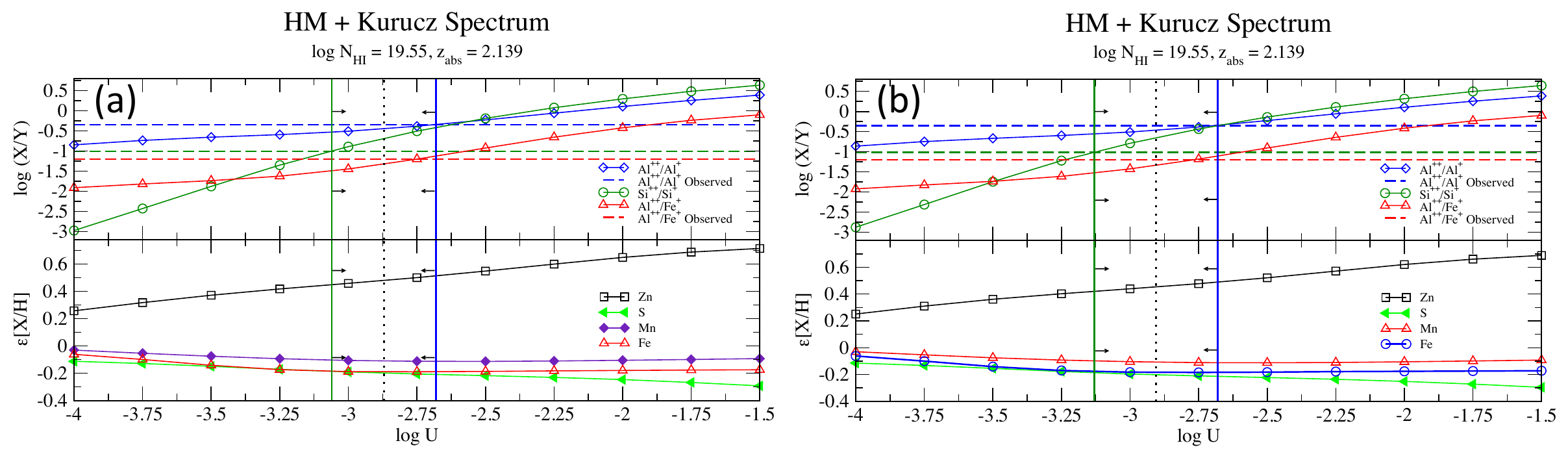}
\caption{
Comparison of the ionization corrections and ionization parameters 
derived for the sub-DLA toward Q1039-2719 with Cloudy version C13.02 
(a) before and (b) after incorporating our revised atomic data for \ion{Zn}{2}.
}
\label{fig_q1039comp}
\end{figure}

\begin{figure}
\includegraphics[scale=0.67]{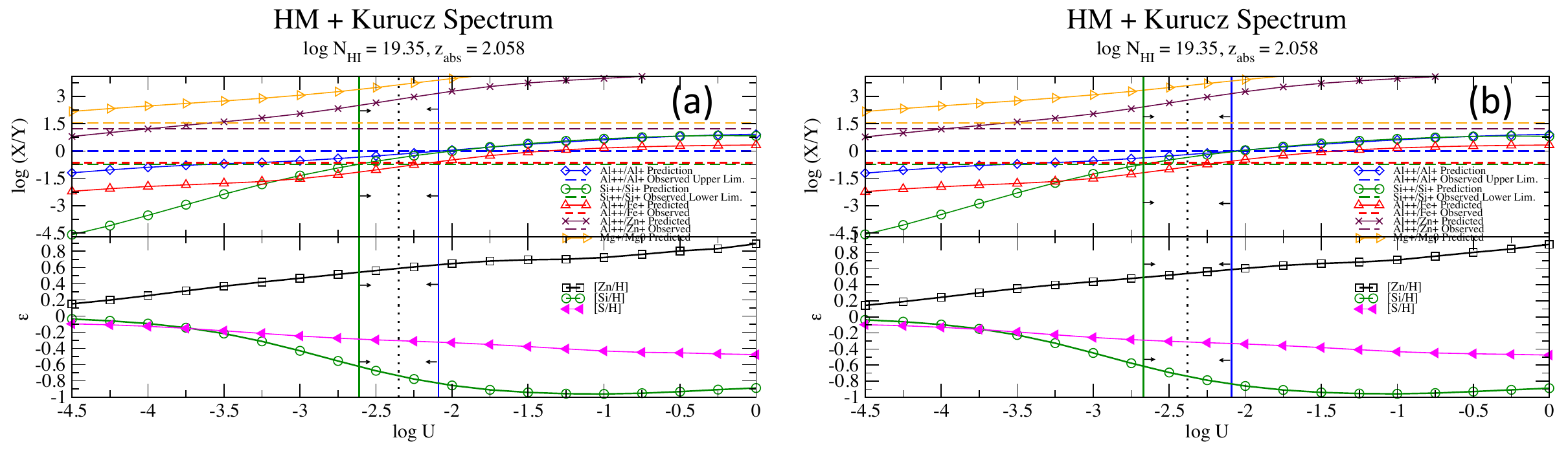}
\caption{ 
Comparison of the ionization corrections and ionization parameters 
derived for the sub-DLA toward Q2123-0050 with Cloudy version C13.02 
(a) before  and (b) after incorporating our revised atomic data for \ion{Zn}{2}. 
}
\label{fig_q2123comp}
\end{figure}

\begin{figure}
\includegraphics[scale=0.67]{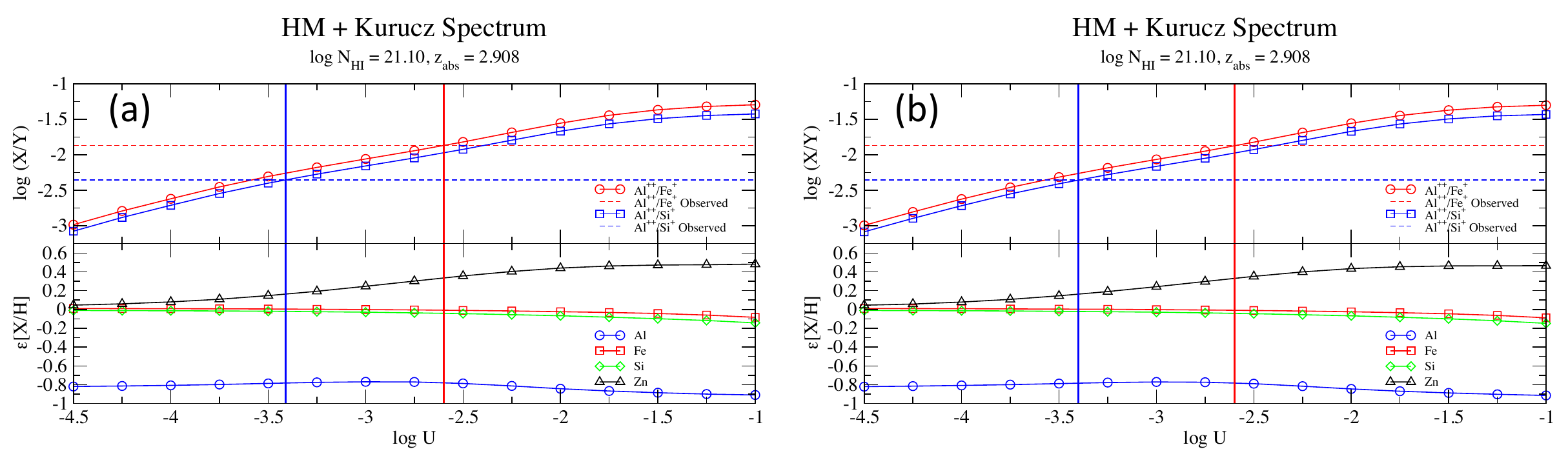}
\caption{ 
Comparison of the ionization corrections and ionization parameters 
derived for the DLA toward Q2342+34 with Cloudy version C13.02 
(a) before  and (b) after incorporating our revised atomic data for \ion{Zn}{2}. 
}
\label{fig_q2342comp}
\end{figure}

Most models of chemical evolution predict a near-solar mean interstellar 
metallicity for galaxies at redshifts $z \sim 0$ [e.g., \citet{Pei99}; 
\citet{Somer01}]. Surprisingly, DLAs at $z< 1.5$ evolve little if at all, with 
metallicities far below the predictions of models based on the cosmic star 
formation history [e.g., \citet{Pet99}; \citet{KF02}; \citet{Pro03, Pro07}; 
\citet{Khare04}; \citet{Kulk05, Kulk07, Kulk10}; \citet{Per08}]. The DLA global 
mean metallicity shows some evolution at high z, but seems to reach only about 
$20 \%$ of the solar level by $z = 0$. These results appear to contradict the 
near-solar mass-weighted mean metallicity at $z \sim 0$ predicted by most models 
and observed in nearby galaxies.  Equally surprising, a significant fraction of 
sub-DLAs appear to be highly metal-rich (near-solar or super-solar), even at 
redshifts $z >2$ (e.g., \citet{Aker05}; \citet{Per06a, Per06b, Per08}, 
\citet{Pro06}; \citet{Mei07, Mei08, Mei09a, Mei09b}; \citet{Kulk07, Kulk10}; 
\citet{Nestor08}; \citet{Som13, Som14}). The super-solar metallicities observed 
in many sub-DLAs (a large fraction of which are derived from Zn) are 
particularly striking, because no local counterparts to such systems are known 
among normal galaxies. 

Given these surprising results, it is natural to ask to what extent the results 
are affected by ionization of the absorbing gas. Ionization corrections are 
likely to be especially important for sub-DLAs and low-$N_{\mathrm{H\,I}}$ DLAs.    
Cloudy calculations using existing atomic data suggest that the 
low-$N_{\mathrm H\,I}$ sub-DLAs can be significantly ionized, but  give relatively 
small ionization corrections to the abundances 
($\varepsilon(\mathrm{X}) = \mathrm{[X/H]_{total} -[X\,II /[H\,I]}$) over the 
range of ionization parameters 
($U = n_{\gamma} /n_{\mathrm{H}}=\Phi_{912} / cn_{\mathrm{H}}$, where 
$\Phi_{912}$ denotes the flux of radiation with h$\nu >$ 13.6 eV and 
$n_{\mathrm{H}}$ denotes the total gas density) allowed by observed ratios such as 
\ion{Al}{3}/\ion{Al}{2}, \ion{S}{3}/\ion{S}{2}, or \ion{Fe}{3}/\ion{Fe}{2} [e.g., 
\citet{DZ03},  \citet{DZ04}; \citet{Mei07,Mei09b}; \citet{Som13, Som14}]. 

To illustrate the effect of ionization, we now show the Cloudy photoionization 
modeling calculations for a few illustrative absorbers. As typical examples of 
sub-DLAs, we choose the $z=2.139$ system toward Q1039-2719 and the $z=2.058$ 
system toward Q2123-0050. Both of these sub-DLAs have log $N_{\rm H I} = 19.35$. 
As an example of DLAs, we use the $z=2.908$ absorber toward Q2342+34 with 
$\log N_{\rm H I} = 21.10$. We 
assume that the ionizing radiation incident on the absorbing cloud is a combination  
of the extragalactic UV background and an O/B-type stellar radiation field. 
We adopt the extragalactic UV background from \citet{HM96} and 
\citet{Madau99}, evaluated at the absorber redshift. The O/B type 
stellar radiation field corresponds to a Kurucz model stellar spectrum for 30,000 K. 
The incident radiation field was taken to be a mixture of the extragalactic and O/B type 
stellar radiation fields in equal parts.  \citet{Schaye06} has suggested that the 
contribution from local sources to the ionization of DLA absorbers may be significant when compared with the contribution from the extragalactic background ionizing radiation. 
Additionally, our simulations include the cosmic microwave background at the absorber redshift, and the cosmic ray background. However, we do not include the 
radiation from local shocks produced by supernovae, white dwarfs, or compact binary systems.

For each absorber, we first ran grids of photoionization models using 
Cloudy version C13.02 [\citet{Ferland13}], by varying the ionization parameter from 
$10^{-6}$ to 1. The models were made to match the observed H I column density 
and the observed metallicity based on \ion{Zn}{2}. Constraints on the ionization parameter were 
estimated by comparing the observed values of the column density ratios 
for various ions with the values calculated from our simulation grids. 
The ionization parameter thus estimated was used to obtain the ionization 
correction values to the abundances. In particular, we used column density 
ratios of adjacent ions of the same element, because they provide more reliable 
observational constraints than the ratios involving different elements, since 
the latter may be affected by differences in dust depletion or in 
nucleosynthesis. For the two illustrative absorbers discussed above toward 
Q1039-2719 and Q2123-0050, we estimate ionization parameters $\log U$ of -2.87 
and -2.35, respectively. The corresponding estimates of ionization corrections 
for \ion{Zn}{2} are 0.48 dex for the absorber toward Q1039-2719 and 0.59 dex for 
the absorber toward Q2123-0050, respectively. For \ion{S}{2}, the corresponding 
ionization corrections are -0.20 dex and -0.30 dex, respectively. 
Figures \ref{fig_q1039comp}a and \ref{fig_q2123comp}a show the ionization 
corrections for several elements as a function of the ionization parameter for 
these models, and the range of $U$ allowed by the observed ion ratios in these 
absorbers. For the DLA toward Q2342+34, we estimate ionization parameter $\log U$
of -3.41 using \ion{Al}{3}/\ion{Si}{2} and \ion{Zn}{2} ionization correction 
of +0.16 dex (Fig. \ref{fig_q2342comp}a).

To assess the effect of our atomic data calculations, we next ran the revised 
Cloudy models that include our improved Zn atomic data for the illustrative DLA 
and sub-DLA absorbers discussed above. Figures \ref{fig_q1039comp}b and 
\ref{fig_q2123comp}b show the results for the ionization correction as a 
function of ionization parameter, and the range of $U$ allowed by the observed 
ion ratios. The ionization parameters are -2.90 and -2.38, respectively, for the 
absorbers toward Q1039-2719 and Q2123-0050. The corresponding ionization 
corrections for \ion{Zn}{2} are 0.45 dex and 0.54 dex, respectively, i.e. lower 
than those obtained with Cloudy version C13.02 by 0.03 dex and 0.05 dex, 
respectively. For the DLA toward Q2342+34, we estimate ionization parameter 
$\log U$ of -3.40 using \ion{Al}{3}/\ion{Si}{2} and \ion{Zn}{2} ionization 
correction of +0.16 dex (Fig. \ref{fig_q2342comp}b).

Combining the difference in ionization correction with the difference in 
$N_{\rm Zn \, II}$ implied by the revised oscillator strengths (which are lower 
by 0.10 dex), the revised [Zn/H] values for the above sub-DLA absorbers would be 
lower by $0.13-0.15$ dex compared to the values based on the previously 
available atomic data. For the DLA absorber, the revised [Zn/H] value would be 
lower by 0.10 dex compared to the values based on the previous atomic data. 

We note, however, that the estimates of ionization corrections are sensitive to 
the adopted values of dielectronic recombination rates, which are unknown for 
ions of most elements in the third row of the periodic table and beyond, 
including key elements such as Al, Fe, Zn. In future papers, we plan to address 
the dielectronic recombination rates. 

\subsection{Emission lines}

\begin{figure}
\includegraphics[scale=0.5]{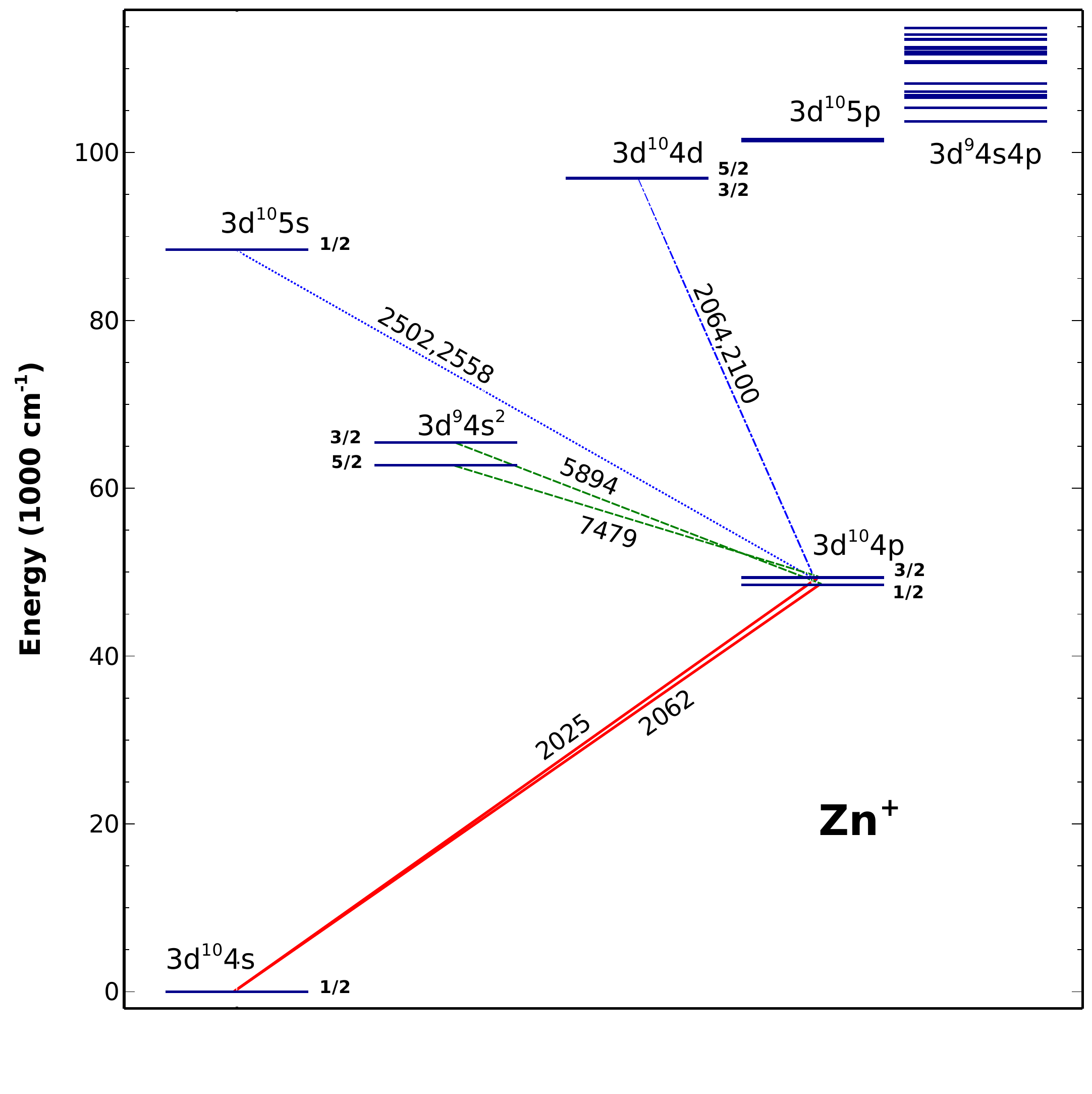}
\caption{ 
The diagram of the lowest levels for \ion{Zn}{2}, showing the even-parity
configuration $3d^{10}4s$, $3d^{10}5s$, $3d^{10}4d$, $3d^94s^2$ levels and the
odd-parity configuration  $3d^{10}4p$, $3d^94s4p$ levels. The strongest lines
originating from the $3d^{10}4p - 3d^{10}4s$ transition (solid lines) and 
from the $3d^{9}4s^2 - 3d^{10}4p$ transition (dashed lines) are also plotted.
Other strong doublet E1 lines, representing transitions $3d^{10}5s - 3d^{10}4p$
(dotted line) and $3d^{10}4d - 3d^{10}4p$ (dash-dotted line) are also shown.
\label{level}
}
\end{figure}

The atomic data described in this paper will be part of the next release of 
Cloudy. For reference, Figure \ref{level} shows the lowest energy levels and 
indicates some of the stronger lines with their air wavelengths given in 
\AA. We know of no calculations of the emission spectra that demonstrates
its diagnostic power, so we do representative calculations here.

\begin{figure}
\includegraphics[scale=0.8]{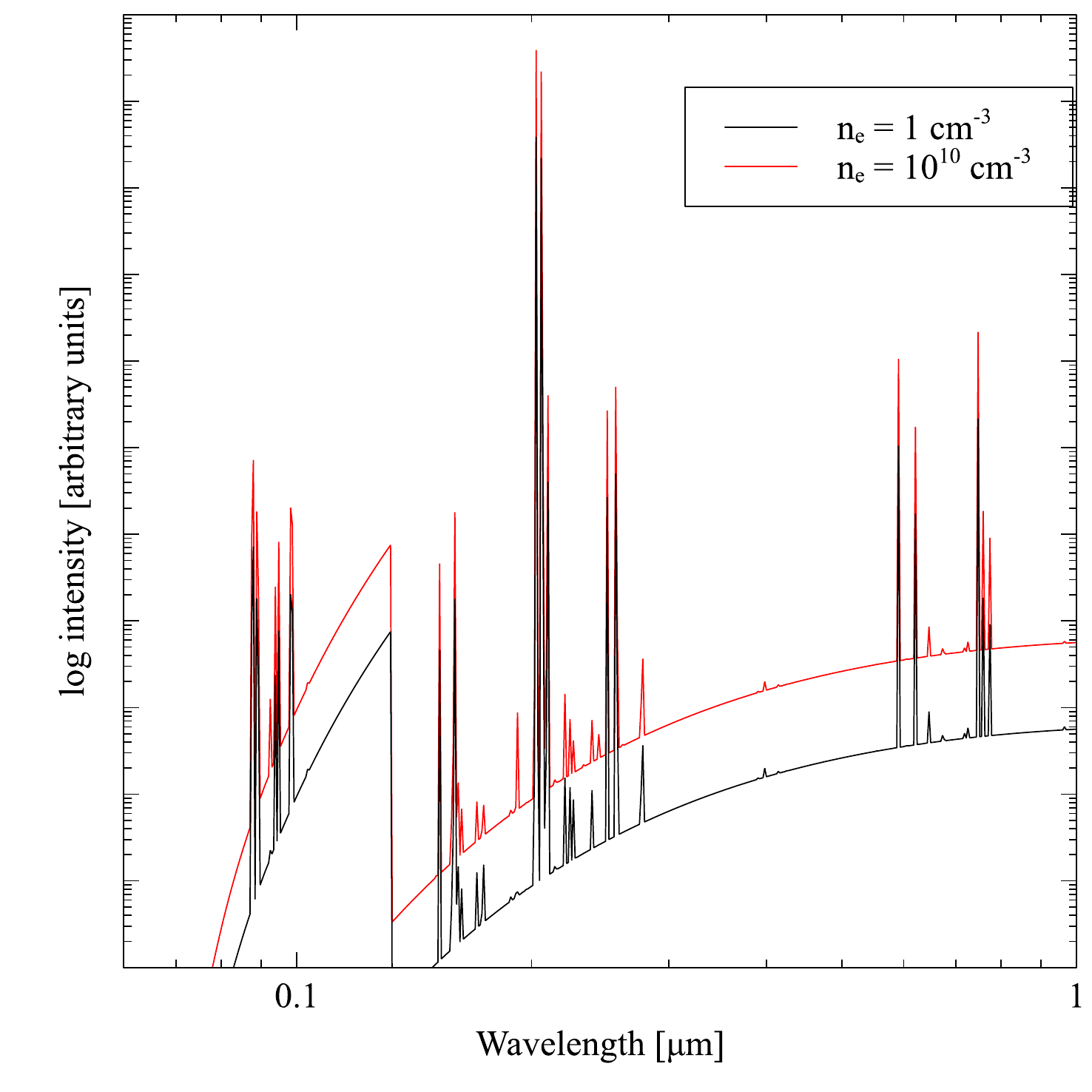}
\caption{ 
The  \ion{Zn}{2} strongest line and continuum emission.
The gas has only Zn$^+$ at 10$^4$~K and is computed at two densities. 
The vertical scale is adjusted so that the two spectra lie near one another,
so the vertical scale is arbitrary.
Both  \ion{Zn}{2}  recombination and brews emission are present.
\label{spectra}
}
\end{figure}

Figure \ref{spectra} shows \ion{Zn}{2} emission spectra at a temperature
of 10$^4$~K and electron densities of 1 and 10$^{10}$ cm$^{-3}$.
The gas was assumed to be composed entirely of Zn$^+$ with no incident SED,
so the emission is entirely due to electron impact excitation.
The strongest lines, as expected, are the 
$3d^{10}4p\;^2P_{3/2,1/2} - 3d^{10}4s\;^2S_{1/2}$ doublet at 
$\lambda \lambda$ 2025.48\AA, 2062.00\AA.
The next stronger lines are 
$3d^{9}4s^2\;^2D_{5/2} - 3d^{10}4p\;^2P_{3/2} $ \, at 7478.82\AA \,
and $3d^{9}4s^2\;^2D_{3/2} - 3d^{10}4p\;^2P_{1/2} $ \,at 5894.37\AA.
These lines are considerably fainter but are important because they can
be detected with large ground-based instruments. The same is true for the E1
doublet lines $3d^{10}5s\;^2S_{1/2} - 3d^{10}4p\;^2P_{1/2,3/2}$ at 
$\lambda \lambda$ 2501.99\AA, 2559.95\AA \, and the doublet of the lines at 
$\lambda \lambda$ 2064.23\AA, 2099.94\AA, representing the E1 transitions  
$3d^{10}4d - 3d^{10}4p$.

\begin{figure}
\includegraphics[scale=0.8]{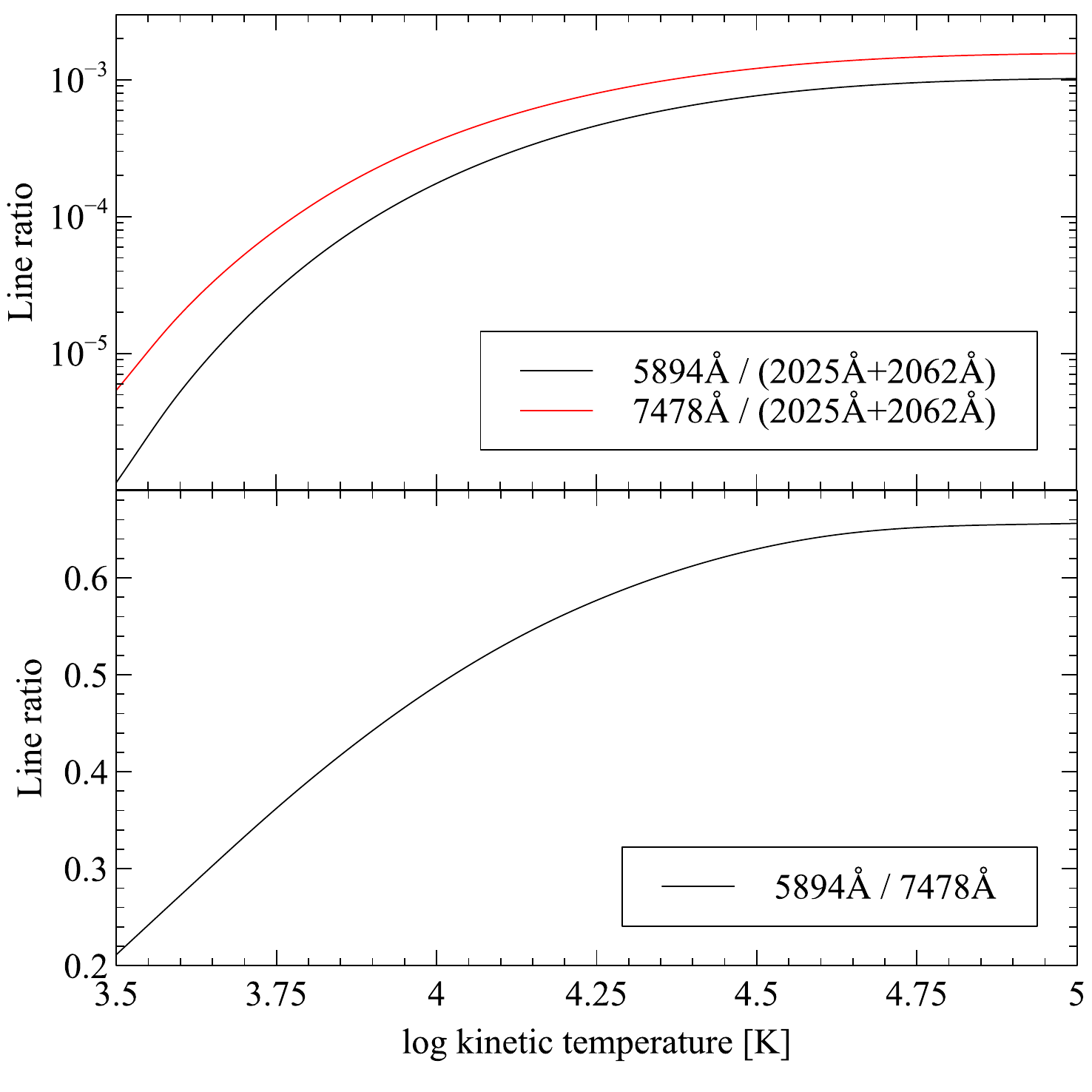}
\caption{ 
Several \ion{Zn}{2} temperature indicators are shown.
\label{Tindicators}
}
\end{figure}

Figure \ref{Tindicators} shows several possible \ion{Zn}{2}  temperature 
indicators. To do this, the kinetic temperature of a pure Zn$^+$ gas with an 
electron density of 1 cm$^{-3}$ was varied over a wide range and the emission 
ratios suggested by Figure \ref{level} plotted.

The upper panel shows the ratio of the two optical lines relative to the sum of 
the two strongest UV lines. This is an indicator with a wide dynamic range 
although the plot also shows that the optical lines are far fainter than the UV 
transitions. The lower panel shows the ratio of the two optical lines. These 
have the advantage of being detectable by ground-based instrumentation.

\section{Conclusions}

Our estimates of the oscillator strengths for the key \ion{Zn}{2} absorption 
lines at 2026.14, 2062.66 {\AA} are higher than the previous values by 0.1 dex, 
implying the \ion{Zn}{2} column densities inferred from these lines to be lower 
by 0.1 dex. Moreover, the sub-DLA ionization corrections for \ion{Zn}{2} would 
be lower by $\sim 0.03-0.05$ dex, as discussed in section \ref{dla}. Thus, the 
logarithmic metallicities inferred from these lines would be lower by 
$0.13-0.15$ dex for sub-DLAs and by 0.1 dex for DLAs, compared to past studies. 
While differences of this amount are significant, they are not adequate to 
explain why the sub-DLA metallicities are so much higher than that of DLAs. 

Using Cloudy simulations, we have demonstrated some astrophysical applications 
of our atomic data calculations, and the predictions for many emission and 
absorption lines. One can compare such predictions with the observed line 
strengths to obtain improved constraints on the chemical composition and 
physical properties of the Galactic and extragalactic ISM. Past observations 
of \ion{Zn}{2} absorption in the Galactic ISM and the DLAs/sub-DLAs have 
targeted the $\lambda \lambda 2026.14, 2062.66$ lines. Our new calculations show 
that these are indeed the strongest observable lines in the commonly accessible 
wavelength region. Our complete set of \ion{Zn}{2} oscillator strengths 
(available online) contains many UV transitions. Although most of these 
transitions are weak, some of them may be detectable in high-S/N spectra 
obtained with the extremely large telescopes of the future. 

\acknowledgments

This work is supported by the collaborative National Science Foundation grants 
AST/1109061 to Univ. of Kentucky and AST/1108830 to Univ. of South Carolina. 
VPK also acknowledges partial support from STScI (HST-GO-12536) and NASA 
(NNX14AG74G). 
GJF acknowledges support by NSF (1108928, 1109061, and 1412155), 
NASA (10-ATP10-0053, 10-ADAP10-0073, NNX12AH73G, and ATP13-0153), 
STScI (HST-AR- 13245, GO-12560, HST-GO-12309, GO-13310.002-A,
and HST-AR-13914), and also is grateful to the Leverhulme Trust for 
support via the award of a Visiting Professorship at Queen's University Belfast.
RK and PB acknowledge support from the European Social Fund under 
the Global Grant measure, project VP1-3.1-{\v S}MM-07-K-02-013.


\begin{thebibliography}{}

\bibitem[Akerman et al.(2005)]{Aker05} 
Akerman, C. J., Ellison, S. L., Pettini, M., \& Steidel, C. C. 
2005, A\&A, 440, 499

\bibitem[Battisti et al.(2012)]{Battisti12} 
Battisti, A. J., et al. 
2012, ApJ, 744, 93

\bibitem[Bergeson \& Lawler(1993)]{berlaw93}
Bergeson, S. D., \& Lawler, J. E.
1993, ApJ, 408, 382

\bibitem[Bogdanovich(2004)]{tro04}
Bogdanovich, P.
2004, Lithuan. J. Phys., 44, 135 

\bibitem[Bogdanovich(2005)]{tro05}
Bogdanovich, P.
2005, Nuclear Instr. Meth. B, 235, 92

\bibitem[Bogdanovich \& Karpu{\v s}kien{\. e}(1999)]{pb99a}
Bogdanovich, P., \& Karpu{\v s}kien{\. e}, R.
1999, Lithuan J. Phys., 39, 193

\bibitem[Bogdanovich \& Karpu{\v s}kien{\. e}(2001)]{pbrk01}
Bogdanovich, P., \& Karpu{\v s}kien{\. e}, R.
2001, Comput. Phys. Commun., 134, 231


\bibitem[Bogdanovich et al.(2002)]{pbrkam02}
Bogdanovich, P., Karpu{\v s}kien{\. e}, R., \& Momkauskait{\. e}, A.
2002,	Comput. Phys. Commun., 143, 174

\bibitem[Bogdanovich et al.(2003a)]{pbrkim03}
Bogdanovich, P., Karpu{\v s}kien{\. e}, R.,\&  Martinson, I.
2003, Phys. Scripta, 67, 44

\bibitem[Bogdanovich et al.(2003b)]{pbrkau03}
Bogdanovich, P., Karpu{\v s}kien{\. e}, R.,\&  Udris, A.
2003, Phys. Scripta, 67, 395

\bibitem[Bogdanovich et al.(2014)]{pbrkds14}
Bogdanovich, P., Kisielius, R., \& Stonys, D.
2014, Lithuan. J Phys., 54, 67

\bibitem[Bogdanovich \& Martinson(1999)]{pbim99}
Bogdanovich, P., \& Martinson, I.
2003, Phys. Scripta, 60, 217

\bibitem[Bogdanovich \& Momkauskait{\. e}(2004)]{pbam04}
Bogdanovich, P.,  \& Momkauskait{\. e}, A.
2004, Comput. Phys. Commun., 157, 217

\bibitem[{\c C}elik et al.(2013)]{celik2013}
{\c C}elik, G., Erol, E., \& Tasfer, M.
2013, JQSRT, 129, 263

\bibitem[Chen et al.(2005)]{Chen05} 
Chen, H.-W., Prochaska, J. X., Bloom, J. S., \& Thompson, I. B. 
2005, ApJ, 634, L25

\bibitem[Chou \& Johnson(1997)]{chou97}
Chou, H.-S., \& Johnson, W. R.
1997, Phys. Rev. A, 56, 2424

\bibitem[Cooke et al.(2011)]{Cooke11} 
Cooke, R., Pettini, M., Steidel, C. C., Rudie, G. C., \& Nissen, P. E. 
2011, MNRAS, 417, 1534

\bibitem[Cowan(1981)]{cowan}
Cowan, R. D. {\em The Theory of Atomic Structure and Spectra}\/ 
University of California Press, Los Angeles, 1981

\bibitem[Dessagues-Zavadsky et al.(2003)]{DZ03} 
Dessauges-Zavadsky, M., P\'eroux, C., Kim, T.-S., D'Odorico, S., \& McMahon, R. G. 
2003, MNRAS, 345,447

\bibitem[Dessagues-Zavadsky et al.(2004)]{DZ04} 
Dessauges-Zavadsky, M., Calura, F., Prochaska, J. X., D'Odorico, S., \& Matteucci, F. 
2004, A\&A, 416, 79


\bibitem[Ferland et al.(2013)]{Ferland13}
Ferland, G. J., Porter, R. L., van Hoof, P. A. M., et al.
2013, Rev. Mexicana Astron. Astrofis., 49, 1  
      

\bibitem[Froese Fischer(1987)]{cff77}
Froese Fischer, C.
1977, J. Phys. B, 10, 1241

\bibitem[Froese Fischer(1987)]{cff87}
Froese Fischer, C.
1987, Comput. Phys. Commun., 43, 355

\bibitem[Froese Fischer \& Godefroid(1991)]{cffmg91}
Froese Fischer, C., \& Godefroid, M.R.
1991, Comput. Phys. Commun., 64, 501

\bibitem[Froese Fischer et al.(1991)]{cffmgah91}
Froese Fischer, C., Godefroid, M.R., \& Hibbert, A.
1991, Comput. Phys. Commun., 64, 486

\bibitem[Fynbo et al.(2009)]{Fynbo09} 
Fynbo, J. P. U, Jakobsson, P., Prochaska, J. X., et al. 
2009, ApJS, 185, 526

\bibitem[Gaigalas et al.(1997)]{gg97}
Gaigalas, G., Rudzikas, Z.R., Froese Fischer, C.
1997, J. Phys. B, 30, 3347

\bibitem[Gaigalas et al.(1998)]{gg98}
Gaigalas, G., Rudzikas, Z.R., Froese Fischer, C.
1998, At. Data Nucl. Data Tables, 70, 1

\bibitem[Guimaraes(2012)]{Guimaraes12} 
Guimaraes, R., Noterdaeme, P., Petitjean, P., et al. 
2012, AJ, 143, 147

\bibitem[Gullberg \& Litz{\' e}n(2000)]{gull00}
Gullberg, D., \& Litz{\' e}n, U.
2000, Phys. Scripta, 61, 652

\bibitem[Haardt \& Madau(1996)]{HM96} 
Haardt, F., \& Madau, P. 
1996, ApJ, 461, 20

\bibitem[Harrison \& Hibbert(2003)]{hh2003}
Harrison, S.A., \& Hibbert, A.
2003, MNRAS, 340, 1279

\bibitem[Hibbert et al.(1991)]{ahrgcff91}
Hibbert, A., Glass, R., Froese Fischer, C.
1991, Comput. Phys. Commun., 64, 445

\bibitem[Howk \& Sembach(1999)]{Howk99} 
Howk, J. C.,  \& Sembach, K. R.
1999, ApJ, 523, L141

\bibitem[Jenkins(2009)]{Jenkins09} 
Jenkins, E. B. 
2009, ApJ, 700, 1299

\bibitem[Jorgenson et al.(2013)]{Jorgenson13} 
Jorgenson, R. A., Murphy, M. T., \& Thompson, R.
2013, MNRAS, 435, 482

\bibitem[Karpu{\v s}kien{\. e} \& Bogdanovich(2001)]{rkpb01_zn}
Karpu{\v s}kien{\. e}, R., \& Bogdanovich, P. 
2001, LithJP, 41, 174

\bibitem[Karpu{\v s}kien{\. e} \& Bogdanovich(2003)]{rkpb03}
Karpu{\v s}kien{\. e}, R., \& Bogdanovich, P. 
2003, J. Phys. B, 36, 2145

\bibitem[Karpu{\v s}kien{\. e} et al.(2013)]{rasa13}
Karpu{\v s}kien{\. e}, R., Bogdanovich, P., \& Kisielius, R.
2013, Phys. Rev. A, 88, 022519

\bibitem[Khare et al.(2004)]{Khare04} 
Khare, P., Kulkarni, V. P., Lauroesch, J. T., York, D. G., Crotts, A. P. S., \& Nakamura, O. 
2004, ApJ, 616, 86 

\bibitem[Kisielius et al.(2014)]{rk2014}
Kisielius, R., Kulkarni V. P., Ferland, G. J., Bogdanovich P., \& Lykins M. L.
2014, ApJ, 780, 76 

\bibitem[Kramida et al.(2014)]{asd13}
Kramida, A., Ralchenko, Y., Reader, J., and NIST ASD Team
2014, {\sl NIST Atomic Spectra Database} (version 5.2),
http://physics.nist.gov/asd National Institute of Standards and 
Technology, Gaithersburg, MD

\bibitem[Kulkarni \& Fall(2002)]{KF02} 
Kulkarni, V. P., \& Fall, S. M. 
2002, ApJ, 580, 732 

\bibitem[Kulkarni et al.(2005)]{Kulk05} 
Kulkarni, V. P., Fall, S. M., Lauroesch, J. T., York, D. G., 
Welty, D. E., Khare, P., Truran, J. W. 
2005, ApJ, 618, 68

\bibitem[Kulkarni et al.(2007)]{Kulk07} 
Kulkarni, V. P., Khare, P., P\'eroux, C., et al.  
2007, ApJ, 661, 88 

\bibitem[Kulkarni et al.(2010)]{Kulk10}  
Kulkarni, V. P.,  Khare, P., Som, D., Meiring, J., York, D. G., 
P\'eroux, C., \& Lauroesch, J. T. 
2010, NewA, 15, 735

\bibitem[Kulkarni et al.(2012)]{Kulk12}  
Kulkarni, V. P., Meiring, J., Som, D., et al. 
2012, ApJ, 749, 176

\bibitem[Kupliauskien{\. e} et al.(2006)]{ak06}
Kupliauskien{\. e} A., Bogdanovich, P., Borovik, A.A., 
Zatsarinny, O.I., Grum-Grzhimailo, A.N., \& Bartschat, K.
2006, J. Phys. B, 39, 591


\bibitem[Lykins et al.(2013)]{Lykins13a} 
Lykins, M. L., Ferland, G. J., Porter, R. L., et al.
2013, MNRAS, 429, 3133

\bibitem[Lykins et al.(2014)]{Lykins13b} 
Lykins, M. L., Ferland, G. J., Kisielius R., et al.
2015, ApJ, submitted


\bibitem[Madau et al.(1999)]{Madau99} 
Madau, P., Haardt, F., \& Rees, M. J. 
1999, ApJ, 514, 648

\bibitem[Mayo et al.(2006)]{mayo2006}
Mayo, R., Ortiz, M., \& Campos, J.
2006, Eur. Phys. J. D, 37, 181

\bibitem[Meiring et al.(2007)]{Mei07} 
Meiring, J. D., Lauroesch, J. T., Kulkarni, V. P., 
P\'eroux, C., Khare, P., York, D. G., \& Crotts, A. P. S.  
2007, MNRAS, 376, 557

\bibitem[Meiring et al.(2008)]{Mei08} 
Meiring, J. D., Kulkarni, V. P., Lauroesch, J. T., 
P\'eroux, C., Khare, P., York, D. G., \& Crotts, A. P. S. 
2008, MNRAS, 384, 1015

\bibitem[Meiring et al.(2009a)]{Mei09a} 
Meiring, J. D., Kulkarni, V. P., Lauroesch, J. T., P\'eroux, C., Khare, P., \& York, D. G. 
2009a, MNRAS, 393, 1513

\bibitem[Meiring et al.(2009b)]{Mei09b} 
Meiring, J. D., Lauroesch, J. T., Kulkarni, V. P., P\'eroux, C., Khare, P., \& York, D. G. 
2009b, MNRAS, 397, 2037

\bibitem[Meyer(1989)]{Mey89} 
Meyer, D. M., Welty, D. E. \& York, D. G. 
1989, ApJ, 343, L37

\bibitem[Morton(2003)]{Morton03} 
Morton, D. C. 
2003, ApJS, 149, 205

\bibitem[Nestor et al.(2008)]{Nestor08} 
Nestor, D. B., Pettini, M., Hewett, P. C., Rao, S., \& Wild, V. 
2008, MNRAS, 390, 1670

\bibitem[Noterdaeme et al.(2012)]{Noter12} 
Noterdaeme, P., Petitjean, P., Carithers, W. C., et al. 
2012, A\&A, 547, L1

\bibitem[Osterbrock \& Ferland(2006)]{AGN3} 
Osterbrock, D. E., \& Ferland, G. J. 
2006, Astrophysics of Gaseous Nebulae \& Active Galactic Nuclei, 2nd Ed. 
(Mill Valley; University Science Press) (AGN3)

\bibitem[Pei et al.(1999)]{Pei99} 
Pei, Y. C., Fall, S. M., \& Hauser, M. G. 
1999, ApJ, 522, 604

\bibitem[P\'eroux et al.(2005)]{Per05} 
P\'eroux, C., Dessauges-Zavadsky, M., D'Odorico, S., et al. 
2005,  MNRAS, 363, 479

\bibitem[P\'eroux et al.(2006a)]{Per06a}
P\'eroux, C., Kulkarni, V. P., Meiring, J., Ferlet, R., 
Khare, P., Lauroesch, J. T., Vladilo, G., \& York, D. G. 
2006a, A\&A, 450, 53

\bibitem[P\'eroux et al.(2006b)]{Per06b} 
P\'eroux, C., Meiring, J. D., Kulkarni, V. P., Ferlet, R., 
Khare, P., Lauroesch, J. T., Vladilo, G., \& York, D. G. 
2006b, MNRAS, 372, 369 

\bibitem[P\'eroux et al.(2008)]{Per08} 
P\'eroux, C., Meiring, J. D., Kulkarni, V. P., et al. 
2008, MNRAS, 386, 2209 

\bibitem[Pettini et al.(1994)]{Pet94} 
Pettini, M., Smith, L. J., Hunstead, R. W., \& King, D. L. 
1994,  ApJ, 426, 79

\bibitem[Pettini et al.(1995)]{Pet95} 
Pettini, M., Lipman, K., \& Hunstead, R. W. 
1995, ApJ, 451, 100

\bibitem[Pettini et al.(1997)]{Pettini97} 
Pettini, M., Smith, L. J., King, D. L., \& Hunstead, R. W. 
1997, ApJ, 486, 665

\bibitem[Pettini et al.(1999)]{Pet99} 
Pettini, M., Ellison, S. L., Steidel, C. C., \& Bowen, D. V. 
1999, ApJ, 510, 576

\bibitem[Pindzola et al.(1991)]{pind91}
Pindzola, M. S., Badnell, N. R., Henry, R. J. W., Griffin, D. C., \& van Wyngaarden, W. L.
1991, Phys. Rev. A, 44, 5628

\bibitem[Prochaska et al.(2003)]{Pro03} 
Prochaska, J. X., Gawiser, E., Wolfe, A. M., Castro, S., \& Djorgovski, S. G. 
2003, ApJ, 595, L9

\bibitem[Prochaska et al.(2005)]{Pro05} 
Prochaska, J. X., Herbert-Fort, S., \& Wolfe, A. M. 
2005, ApJ, 635, 123

\bibitem[Prochaska et al.(2006)]{Pro06} 
Prochaska, J. X., O'Meara, J. M., Herbert-Fort, S., Burles, S., Prochter, G. E., \& Bernstein, R. A. 
2006, ApJ, 648, L97

\bibitem[Prochaska et al.(2007)]{Pro07} 
Prochaska, J. X., Chen, H.-W., Dessauges-Zavadsky, M., et al. 
2007a, ApJ, 666, 267

\bibitem[Rafelski et al.(2012)]{Rafelski12} 
Rafelski, M., Wolfe, A. M., Prochaska, J. X., Neeleman, M., \& Mendez, A. J. 
2012, ApJ, 755, 89

\bibitem[Savage \& Sembach(1996)]{Savage96} 
Savage, B. D., \& Sembach, K. R. 
1996, ARAA, 34, 279

\bibitem[Savaglio et al.(2003)]{Savaglio03} 
Savaglio, S., Fall, S. M., \& Fiore, F. 
2003, ApJ, 585, 638

\bibitem[Schaye(2006)]{Schaye06} 
Schaye, J. 
2006, ApJ, 643, 59

\bibitem[Sharma et al.(2011)]{sharma11}
Sharma, I., Surzhykov A., Srivastava, R., \& Fritzsche, S.
2011, Phys. Rev. A, 83, 062701

\bibitem[Som et al.(2013)]{Som13} 
Som, D., Kulkarni, V. P., Meiring, J., et al. 
2013, MNRAS, 435, 1469

\bibitem[Som et al.(2014)]{Som14} 
Som, D., Kulkarni, V. P., Meiring, J., et al. 
2015, ApJ, submitted

\bibitem[Somerville et al.(2001)]{Somer01} 
Somerville, R. S., Primack, J. R., \& Faber, S. M. 
2001, MNRAS, 320, 504

\bibitem[Storrie-Lombardi \& Wolfe(2000)]{Storrie00} 
Storrie-Lombardi, L. J., \& Wolfe, A. M. 
2000, ApJ, 543, 552


\bibitem[Zafar et al.(2013)]{Zafar13}
Zafar, T., P\'eroux, C., Popping, A., et al. 2013,  A\&A, 556, 141

\bibitem[Zatsarinny \& Bandurina(1999)]{oleg99}
Zatsarinny, O., \& Bandurina, L.
1999, J. Phys. B, 32, 4793

\end{thebibliography}
\end{document}